\documentclass[aps,prl,groupedaddress,twocolumn,showpacs,preprintnumbers,amsmath,amssymb]{revtex4}

\usepackage{graphicx}
\usepackage{bm}
\usepackage{eucal}
\usepackage{ulem}
\usepackage{color}
\usepackage{afterpage}

\begin{document}

\title{Morphotropic Phase Boundaries in Ferromagnets: Tb$_{1-x}$Dy$_x$Fe$_2$ Alloys}
\author{Richard Bergstrom Jr.,$^1$ Manfred Wuttig$^1$, James Cullen,$^1$ Peter~Zavalij,$^2$ Robert Briber,$^1$
Cindi Dennis,$^3$ V.\,Ovidiu Garlea,$^4$ and Mark Laver,$^{1,5\mbox{-}7}$
}
\affiliation{
$^1$Dept.\ of Materials Science and Engineering, University of Maryland, College Park, Maryland 20742, USA\\
$^2$Dept.\ of Chemistry and Biochemistry, University of Maryland, College Park, Maryland 20742, USA\\
$^3$National Institute of Standards and Technology, Gaithersburg, Maryland 20899, USA\\
$^4$Quantum Condensed Matter Division, Oak Ridge National Laboratory, Oak Ridge, Tennessee 37881, USA\\
$^5$Laboratory for Neutron Scattering, Paul Scherrer Institut, CH-5232 Villigen PSI, Switzerland\\
$^6$Department of Physics, Technical University of Denmark, DK-2800 Kongens Lyngby, Denmark\\
$^7$Niels Bohr Institute, University of Copenhagen, DK-2100 Copenhagen, Denmark
}

\begin{abstract}
The structure and properties of the ferromagnet Tb$_{1-x}$Dy$_x$Fe$_2$ are explored through the morphotropic phase boundary (MPB)
separating ferroic phases of differing symmetry.
Our synchrotron data support a first order structural transition, with a broadening MPB width at higher temperatures.
The optimal point for magnetomechanical applications is not centered on the MPB but lies on the rhombohedral side,
where the high striction of the rhombohedral majority phase combines with the softened anisotropy of the MPB.
We compare our findings with single ion crystal field theory and with ferroelectric MPBs,
where the controlling energies are different.
\end{abstract}


\date{\today}
\maketitle

The functional response of ferroelectrics with compositions in the vicinity of a morphotropic phase boundary (MPB) is significantly enhanced.
At the MPB composition PbZr$_{.52}$Ti$_{.48}$O$_3$, lead zirconate-titanate (PZT)
has long been a material of choice for actuator applications~\cite{Jaffe1954}.
New Pb-free piezoelectrics such as $(1-y)$Ba(Zr$_{.2}$Ti$_{.8}$)O$_3$--$y$(Ba$_{.7}$Ca$_{.3}$)TiO$_3$
also utilize the enhanced piezoelectric response near the MPB~\cite{Ehmke2012andLiu2009}.
The MPB separates two phases of distinct symmetry, typically rhombohedral and tetragonal,
resulting in elastic boundaries between phases.
PZT, for example, is a pseudo-binary alloy of tetragonal PbTiO$_3$ and rhombohedral PbZrO$_3$
with spontaneous ferroelectric polarizations oriented along $[100]$ and $[111]$ respectively in the
two phases~\cite{Jaffe1971etc,Noheda2000andSolanki2011,Schonau2007}.
At the atomic scale, two scenarios describing the ferroelectric MPB have been proposed.
Both scenarios try to minimize the elastic and electric depolarization energies,
and appear very similar when probed by bulk diffraction techniques~\cite{Schonau2007,Wang2006and2007}.
In the first, supercells incorporating local monoclinic distortions form along the MPB,
as suggested by synchrotron X-ray diffraction (XRD) from PZT showing
peak splittings typical of a low crystal symmetry~\cite{Noheda2000andSolanki2011}.
In the second, adaptive nanodomains form from orientational variants of the parent phases.
This scenario explains the relations observed between the lattice parameters of the component
phases~\cite{Wang2006and2007,Rossetti2008andKhachaturyan2010},
and is supported by recent high-resolution microscopy studies~\cite{Borisevich2012etc}.

There are at least three types of MPBs distinguished by ferroic order: ferroelectric, ferromagnetic, and ferroelastic.
Hybrid MPBs between ferroics of different nature should also exist~\cite{KhachaturyanTBP}.
Information as to intrinsic MPB behavior can be revealed from studies of ferromagnetic MPBs,
since in ferromagnets, the ferroic order and local atomic displacements are readily separated.
A previous XRD study examined the MPB in the Tb$_{1-y}$Dy$_y$Co$_2$ ferromagnet~\cite{Yang2010}.
As in ferroelectrics, in Tb$_{1-y}$Dy$_y$Co$_2$ the MPB separates rhombohedral and tetragonal structures of the parent compounds TbCo$_2$ and DyCo$_2$,
and coincides with an enhancement in a `figure of merit' of magnetoelastic properties.

\begin{figure}
\includegraphics[width=3.2in]{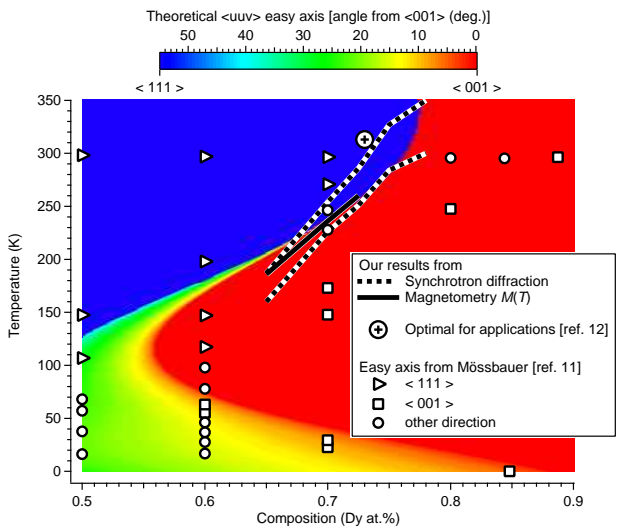}
\caption{(color online) Phase diagram of Tb$_{1-x}$Dy$_x$Fe$_2$ (TDF).
The background shading shows the magnetic easy-axis direction calculated using anisotropy parameters from crystal field theory (see text).
The easy axis reorients from $ \langle 111 \rangle$ for TbFe$_2$ to $\langle 001 \rangle$ for DyFe$_2$.
Overlayed is the morphotropic phase boundary determined from our synchrotron XRD (dotted lines)
and magnetometry (solid line) measurements,
as well as the easy axes reported previously from M{\"o}ssbauer spectroscopy (open symbols)~\cite{Atzmony1977}. 
The cross in a circle indicates the optimal temperature (40\,$^\circ$C) for magnetomechanical device applications,
as determined for TDF $x=.73$~\cite{Clark1985}.}
\end{figure}
\begin{figure*}
\center{\includegraphics[height=3in]{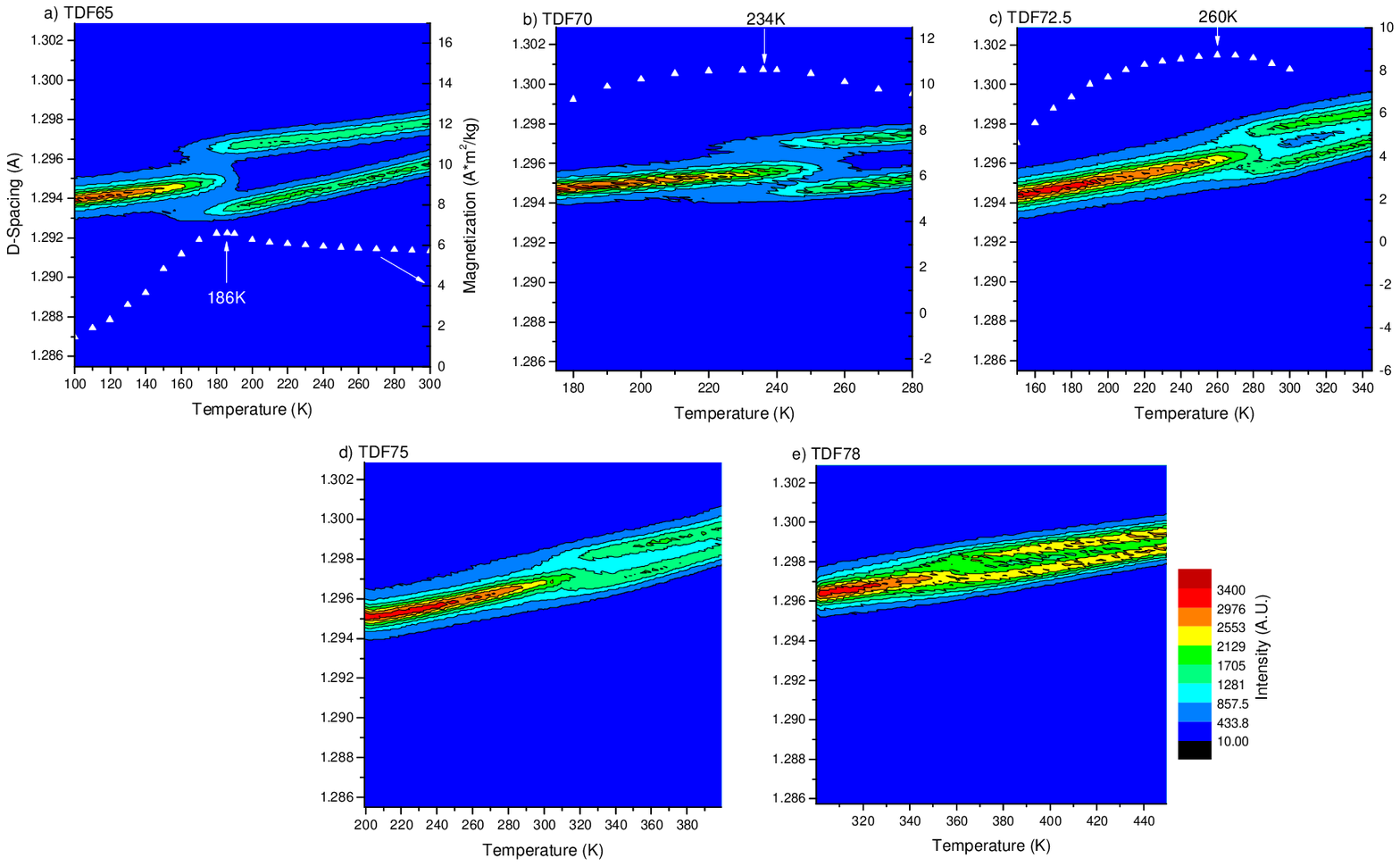}}
\caption{(color online) Contour plots showing the temperature dependence of the diffracted synchrotron X-ray intensity in the vicinity of the
$4\,4\,0$ Bragg reflection across a series of Tb$_{1-x}$Dy$_x$Fe$_2$ compositions
(a) $x = .65$, (b) $x = .70$, (c) $x = .725$, (d) $x=.75$ and (e) $x=.78$.
The solid white triangles in (a), (b) and (c) depict the sample magnetization $M$ in an applied field of 7.96\,kAm$^{-1}$.
Each $M(T)$ curve has its maximum highlighted by an arrow.}
\end{figure*}
Here we report on magnetometry, neutron diffraction and synchrotron XRD experiments on the pseudo-binary alloy Tb$_{1-x}$Dy$_x$Fe$_2$ (TDF).
We show that the ferromagnetic MPB consists of a coexistence of the two crystallographic structures of the parent compounds TbFe$_2$ and DyFe$_2$.
The volume fractions of these components vary continuously across the boundary.
Our measurements are summarized in the phase diagram of Fig.~1 and compared with previous measurements~\cite{Atzmony1977,Clark1985}
plus the results of single ion crystal field theory.
We find that the MPB region, across which phases coexist, widens with increasing temperature.
This reflects entropy-driven microstructural changes that lie largely beyond the scope of single ion (mean field) theory.

Alloys of Tb$_{1-x}$Dy$_x$Fe$_2$ (TDF) with $x=.650$, .675, .700, .725, .750, .765, .780 and .800 at.\,\% were prepared at the
Metals Preparation Center at Ames Laboratory by arc melting the constituent elements on a water-cooled copper hearth plate
in a high-purity argon atmosphere.
To ensure homogeneity each alloy was melted three times before being ground into a powder in an argon atmosphere.
The starting metals were Ames Lab 99.99\% Dy and Tb, and 99.95\% pure electrolytic Fe.
Magnetization was measured using a Quantum Design SQUID.

Zero-field synchrotron XRD experiments were performed at Argonne National Laboratory beamline 11\mbox{-}BM.
The samples for XRD measurements were further ball-milled in an argon atmosphere
before being sealed in 0.3\,mm diameter quartz capillaries.
During the XRD measurement, the sample temperature was controlled by a Cryostream N$_2$ gas blower (range 100\,K to 450\,K)
and the sample rotated continuously to reduce preferred orientation effects.
Twelve silicon (111) crystal analyzers positioned in front of the LaCl$_3$ scintillation detectors
facilitated an instrument resolution of $\Delta d / d \sim 2 \times 10^{-4}$.
Rietveld refinements of XRD data were performed using GSAS~\cite{XRDrefinement}.
Neutron time-of-flight diffraction profiles in the range 12\,K to 320\,K were collected at the POWGEN instrument
at the Spallation Neutron Source at Oak Ridge National Laboratory.
For these measurements, samples were loaded into 8\,mm dia.\ vanadium cans,
and the data refined using the FullProf package~\cite{neutronrefinement}. 

\begin{figure}
\begin{flushleft}
\includegraphics[height=1.5in]{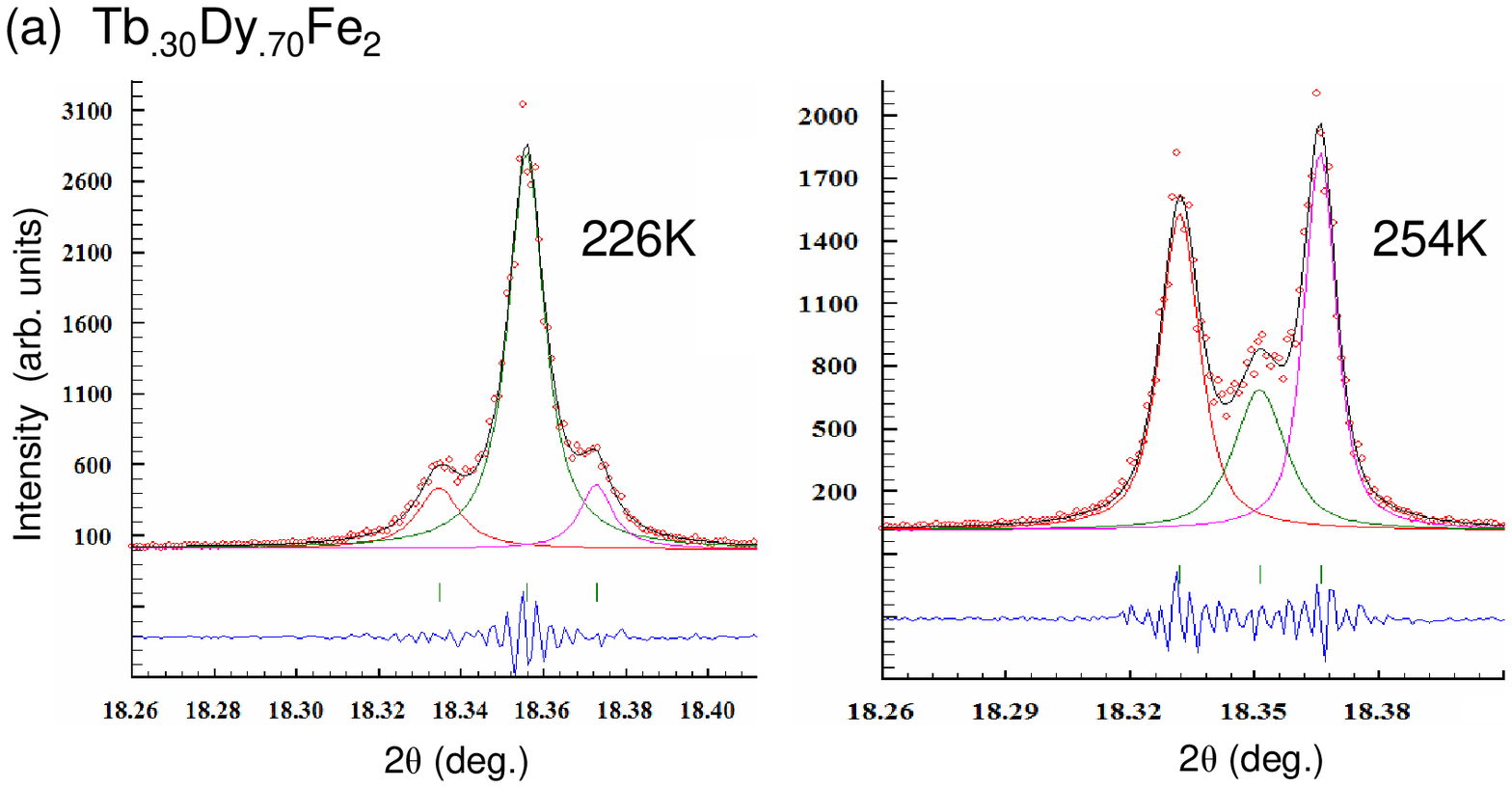}
\end{flushleft}
\center{\includegraphics[width=2.8in]{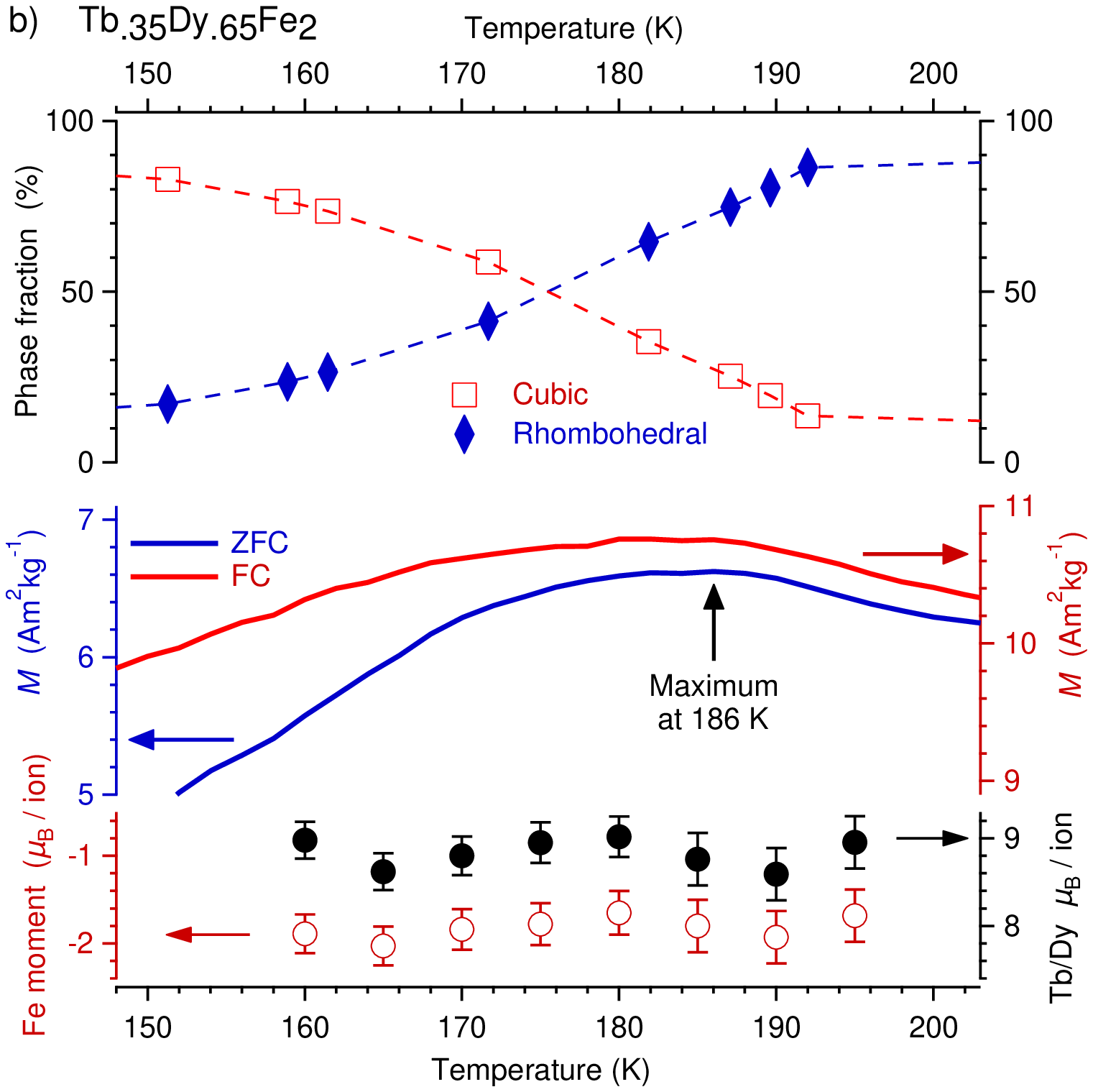}}
\caption{(color online) (a) Typical profile fits of Tb$_{1-x}$Dy$_x$Fe$_2$ synchrotron XRD data.
Here profiles for $x = .70$ in the vicinity of the cubic $4\,4\,0$ Bragg reflection are shown, at temperatures where two structures coexist:
the left (right) panel reveals a 25\% (75\%) rhombohedral phase fraction from Gaussian peak fits (red and green lines).
The blue lines underneath shows the difference plots.
(b) Temperature dependences across the MPB for $x = .65$: (top panel) phase fractions from the fitted peak intensities;
(middle panel) magnetization $M(T)$ measured in an applied field of 7.96\,kAm$^{-1}$;
(bottom panel) ordered atomic moment determined by neutron diffraction.
Dashed lines are guides to the eye.
}
\end{figure}
In Fig.~2 we show XRD data as a function of temperature across a series of Tb$_{1-x}$Dy$_x$Fe$_2$ (TDF) alloys with $x$ in the range .65 to .78 at.\,\%.
We focus on the $4\,4\,0$ Bragg reflection (cubic or pseudo-cubic notation is used throughout this Letter),
compiling contour plots from the powder XRD patterns taken in steps of $\approx 2.5$\,K.
Across the entire TDF series, these contour plots (Fig.~2) show wide temperature regions over which two phases coexist.
At high temperatures, the XRD pattern can be fitted with rhombohedral R$\bar{3}$m symmetry~\cite{Yang2008}.
The low temperature phase was anticipated to be tetragonal I4$_1/$amd, but is fitted with cubic Fd$\bar{3}$m symmetry,
as the synchrotron XRD instrument resolution is unable to resolve the small tetragonal distortion $\lesssim 1 \times 10^{-4}$
even in the parent DyFe$_2$~\cite{Barbara1977}.
In the transition region, the $4\,4\,0$ reflection is clearly split into three [Fig.~3(a)]
corresponding to a coexistence of two peaks from the high-temperature rhombohedral phase and a single peak from the low-temperature phase.
Systematic profile fitting reveals that the width of the two-phase region increases
as the DyFe$_2$ concentration $x$ increases, or equivalently as the MPB transition temperature increases.
This is summarized graphically in the phase diagram (Fig.~1)
where dotted lines delineate the two-phase region in which both phase fractions are $> 25$\%.

Rietveld refinement of synchrotron powder XRD patterns has become a standard tool for exploring the structural changes around the
MPB~\cite{Schonau2007,Noheda2000andSolanki2011}.
Motivated by the reports of an intermediate monoclinic phase in ferroelectrics~\cite{Noheda2000andSolanki2011},
a third phase was trialed in the MPB region of TDF.
For TDF65 at 175\,K, for example, we found that the goodness-of-fit and Rietveld indices $\chi^2$ and $R_\textrm{Bragg}$
improve to some extent from $\chi^2 = 1.87$ to 1.25 and from $R_\textrm{Bragg} = 0.0685$ to 0.0509
if a second cubic phase with a different lattice constant is added.
The refined strains of all the phases are then reduced.
On inspection, however, the peaks do not appear visibly split in support of a third phase [c.f.\ Fig.~3(a)].
We surmise that two highly-strained phases coexist across the MPB.
The relative populations of the two phases as measured by the integrated diffracted intensities
are plotted for TDF65 in Fig.~3(b) along with our magnetometry and neutron diffraction results.
The magnetization $M(T)$ measured in a low applied field (here we show results for 7.96\,kAm$^{-1}$)
shows a peak at 186\,K after both zero field-cooling (ZFC) and field-cooling.
We shall see that this peak corresponds to a reduced magnetic anisotropy at the MPB.
Our $M(T)$ data for TDF70 and TDF72.5 reveal peaks at 236\,K and 260\,K respectively,
as highlighted for the ZFC data overlayed on Fig.~2.
Like the structural coexistence region, the maximum in $M(T)$ broadens as the MPB transition temperature increases.
We note that analogous behavior has not been reported in ferroelectrics.

A neutron powder diffraction study was carried out to probe possible changes in the local magnetization.
The resolution of the neutron instrument is lower than that of a synchrotron diffractometer
and proved insufficient to distinguish either the directions of the magnetic moments or the rhombohedral splitting.
All neutron data were accordingly refined in a cubic setting.
Profiles were collected over the MPB in TDF65.
In the refinements, thermal broadening parameters coupling to the peak widths were found to climb rapidly with increasing temperature
as the MPB region is entered,
before resuming a normal slow rise with temperature above the coexistence region.
This is consistent with enhanced accommodation strain in the MPB region.
Magnetic moments were refined for the rare earth and iron sites simultaneously,
yielding values of $\approx 9$\,$\mu_\mathrm{B}$ at the rare earth site and $\approx -1.9$\,$\mu_\mathrm{B}$ for the iron.
These remain invariant within experimental error across the MPB [Fig.~3(b)].

In Fig.~1 we compare the phase boundary determined from our synchrotron XRD and DC magnetometry measurements
with previous results on the magnetic moment directions deduced from M\"ossbauer spectra~\cite{Atzmony1977}.
The agreement between the three datasets is excellent.
It is revealing to see to what extent these observations can be explained by single ion
crystal field theory~\cite{Callen1965and1966,Koon1978,Martin2006,Bowden2004}.
In Fig.~1 we plot, as background shading, the magnetic easy-axis direction determined by minimizing the free energy
consisting of magnetocrystalline anisotropy terms $K_1$, $K_2$, $K_3$, $K_4$,
including a phenomenological correction $\Delta K_1$ to model the leading-order magnetostrictive effects~\cite{Koon1978}.
The magnetostriction and $K_i$'s are temperature dependent, and are presumed to vary linearly with composition, following relations of the form
$K_i(\mathrm{alloy}) = (1-x) K_i(\mathrm{TbFe}_2) + x K_i(\mathrm{DyFe}_2)$.
The $K_i$'s for the parent compounds TbFe$_2$ and DyFe$_2$ were calculated in Ref.~\onlinecite{Martin2006}.
At temperatures $\approx 300$\,K, the observed location of the MPB is coincident with the model predictions (Fig.~1),
though elsewhere the agreement is not as good.

\begin{figure}
\includegraphics[width=3.2in]{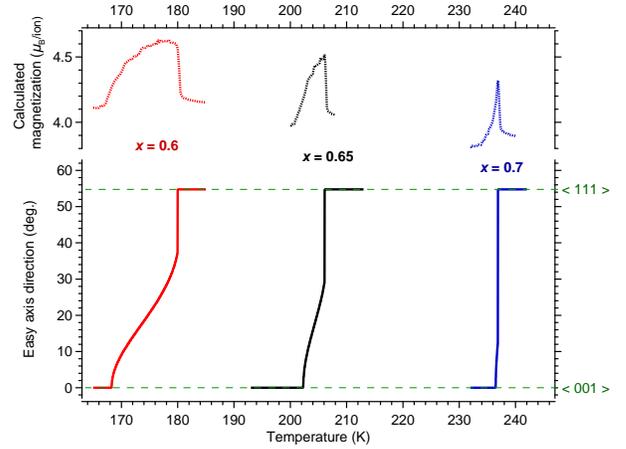}
\caption{(color online) Theoretical behavior across the phase boundary using the single ion crystal field model (see text).
Selected Tb$_{1-x}$Dy$_x$Fe$_2$ compositions are shown.
The calculated easy-axis directions are found to be of $\langle uuv \rangle$-type;
the solid lines (bottom panel) describe the angle of the easy axis from $\langle 001 \rangle$.
The dotted lines (top panel) show the calculated magnetization at 7.96\,kAm$^{-1}$. 
}
\end{figure}
The single ion model explains the extended nature of the transition in easy axis from $\langle 001 \rangle$ to $\langle 111 \rangle$ at the MPB.
However we will see that the temperature dependence of the transition width is not accounted for.
In the model, the spin reorientation occurs in two steps as illustrated by cuts at constant composition (Fig.~4).
Throughout, the easy axis is found to be of $\langle uuv \rangle$-type.
The first step is a second-order transition out of the $\langle 001 \rangle$ phase
with the easy axis rotating continuously with either temperature or composition.
This step is linked to the first-order magnetocrystalline term $(K_1 + \Delta K_1)$ transiting from positive to negative.
Continuous reorientations are permitted since $K_3$ and $K_4$ are non-zero,
accounting for the easy-axis directions away from principal crystal axes observed at low temperatures~\cite{Atzmony1977,Atzmony1976}.
The second step in the reorientation is a first-order transition to $\langle 111 \rangle$, exhibiting a discontinuity in the easy-axis direction.

The softening anisotropy at the phase boundary is anticipated to yield a maximum in the susceptibility,
and indeed this is seen by adding a $\mathbf{M}.\mathbf{H}$ term in the free energy that couples to the applied field $\mathbf{H}$.
The calculated polycrystalline $M(T)$ is maximal at the first-order transition rather than at the second-order transition (Fig.~4),
consistent with the proposition that a discontinuity in the ferroic order at the MPB is prerequisite for enhanced properties~\cite{LiuLi2011}.
This can be borne in mind when comparing the experimental $M(T)$ with the structural data [Fig.~3(b)],
in particular noting that the optimal point for magnetomechanical device applications lies in the higher temperature part of the MPB region
(Fig.~1)~\cite{Clark1985}.
In this part of the phase diagram, two ideal effects coincide: the high striction of the rhombohedral majority phase combined with the
softened anisotropy of the MPB region.

With increasing MPB temperature, the single ion model predicts a diminishing width of the reorientation region
following the shrinking of the second-order region (Fig.~4).
This is in sharp contrast with experiment, where an increasing width of the structural coexistence region is observed.
We find that atomic diffusion, which might be anticipated to be significant at elevated temperatures,
provides a negligible contribution in accounting for the broadened MPB character.
At 500\,K in TDF the Fe diffusion rate $\approx 2 \times 10^{-31}$\,cm$^2$s$^{-1}$
based on the reported Co diffusion in Co$_{69}$Nb$_{31}$ which also exhibits a C15 Laves phase structure~\cite{Denkinger2000}.
Taking a scenario of 3 hours at 500\,K,
the average diffusion distance $\approx 5 \times 10^{-7}$\,nm,
a negligible value compared to the TDF lattice constant of 0.73\,nm.

To understand the differences between MPBs in ferromagnets and ferroelectrics,
one must consider the governing energies.
The MPBs in ferroelectrics are thought to consist of nano-twinned meso-structures in which the polarization changes continuously as a function
of the twin concentration as the boundary is traversed~\cite{Wang2006and2007,Rossetti2008andKhachaturyan2010}.
In other words, their structure is elastically controlled and not expected to be strongly temperature dependent.
For MPBs in ferromagnets the situation is rather different.
Here the magnetic polarization is determined by a balance of exchange and magnetocrystalline anisotropy energies.
The latter has two components: atomic shape (crystal field) and magnetoelastic anisotropies.
The atomic shape component $\sim [M(T)/M(0)]^{10}$ from single ion theory~\cite{Callen1965and1966} and dominates at low temperatures.
The magnetoelastic component $\Delta K_1(T)$ can be estimated by single ion theory,
yielding reasonable descriptions of the temperature dependences observed in TbFe$_2$~\cite{Clark1977}, DyFe$_2$~\cite{Bowden2004},
and in TDF (see e.g.\ Ref.~\onlinecite{Clark1985} for $x=.73$).
$\Delta K_1(T)$ is seen to fall off more slowly with temperature $\approx [M(T)/M(0)]^6$~\cite{Callen1965and1966,Clark1985,Clark1977},
becoming dominant at high temperatures.
With these contrasting dependences in mind,
it is not surprising that the MPB behavior is temperature dependent.
In terms of a free energy functional of the magnetic easy-axis direction,
the wider phase coexistence region can be understood
in terms of a flatter energy landscape at higher MPB temperatures~\cite{CullenUnpub}.
There the magnetocrystalline anisotropies $K_i$ evanesce such that the free energy wells of the $\langle 100 \rangle$ and $\langle 111 \rangle$
easy axes become more shallow;
orientational entropy will then co-populate these minima over a broader thermal region.

In conclusion, we have performed magnetometry, synchrotron XRD and neutron diffraction on
Tb$_{1-x}$Dy$_x$Fe$_2$ alloys, focusing on the MPB region in the temperature range 200\,K to 350\,K pertinent to applications.
We have compared our observations with single ion crystal field theory.
The latter indicates that there are two distinct transition regions:
the first, at lower temperatures, consists of a continuous rotation of the magnetic easy axis stabilized by higher order anisotropy terms $K_3$ and $K_4$.
The second is a first-order region that increasingly dominates the MPB at elevated temperatures (Fig.~4).
Our synchrotron XRD profiles are consistent with a first-order transition with continously varying volume fractions between
two structural phases.
The coexistence region broadens as the MPB moves to higher temperatures (Fig.~1).
This contrasts with expectations from single ion theory
and furthermore with the MPB character in ferroelectrics.
The high-temperature broadening of ferromagnetic MPBs can be understood
in terms of diminishing magnetocrystalline anisotropies and associated entropic effects.
More advanced theory and experimental studies are called for to complete the microstructural picture.

The authors would like to thank Denis Sheptyakov and Lukas Keller for assistance in obtaining preliminary diffraction patterns
at the Swiss Spallation Neutron Source SINQ.
ML acknowledges support from DanScatt.
Use of the Advanced Photon Source at Argonne National Laboratory was supported by the U.S. Department of Energy, Office of Science,
Office of Basic Energy Sciences, under Contract No. DE-AC02-06CH11357.
A portion of this research at ORNL's Spallation Neutron Source was sponsored by the Scientific User Facilities Division,
Office of Basic Energy Sciences, U.S. Department of Energy.
This work was sponsored by NSF under grant DMR D~1206397 and benefited from DOE grant DESC0005448.

\end{document}